\documentclass[a4paper]{jpconf}
\usepackage{graphicx}
\begin{document}

\title{Looking into the heart of a beast: the black hole binary LS~5039}

\author{T. Szalai$^1$, L. L. Kiss$^{2,3}$ and G. E. Sarty$^4$}

\address{$^1$Department of Optics and Quantum Electronics, University of
Szeged, D\'om t\'er 9., Szeged H-6720, Hungary}
\address{$^2$Konkoly Observatory of the Hungarian Academy of Sciences, H-1525 Budapest, P.O. Box 67, Hungary}
\address{$^3$Sydney Institute for Astronomy, School of Physics A28, 
University of Sydney, NSW 2006, Australia}
\address{$^4$Department of Physics and Engineering Physics, University of Saskatchewan,
Saskatoon, Saskatchewan S7N 5E2, Canada}

\ead{szaszi@titan.physx.u-szeged.hu}

\begin{abstract}
LS~5039 is a relatively close microquasar consisting of a late O-type star and a compact object (very possibly a black hole)
on a highly eccentric orbit with a period of 3.9 days. The high X-ray, gamma-ray and radio luminosity indicate light-matter 
interaction, which arise from the stellar wind of the primary star accreting toward the black hole. Former examinations 
suggest that LS~5039 could be a prototype of wind-fed high mass X-ray binaries (WXBs) with diskless main sequence O 
primaries.
Now there is a great chance to better understand the configuration and the physical processes in the exotic system. In July 
2009 LS~5039 was followed by the Canadian MOST space telescope to get ultraprecise photometric data in a month-long 
semi-continuous time series. Parallel to this, we have taken simultaneous high-resolution optical spectra using the 2.3m ANU 
telescope of the Siding Spring Observatory, supplemented with further data obtained in early August 2009 with the same 
instrument. 
Here we present the first results from the new echelle spectra, which represent the best optical spectroscopy ever obtained 
for this intriguing system. We determined fundamental orbital and physical parameters of LS~5039 and examined the 
configuration and the circumstellar environment of the system via radial velocity measurements and detailed line-profile 
analysis of H-Balmer, He I and He II lines. 
\end{abstract}

\section{Introduction}

LS~5039 (V479~Sct), classified as a high-mass X-ray binary (HMXB) by Motch et al. (1997), is one of the most
studied members of its class. Based on VLA data, Mart\'i et al. (1998) discovered a
persistent non-thermal radio counterpart, associated with mildly relativistic radio jets (Paredes 
et al. 2000, 2002). In addition, a very high energy (VHE) 
$\gamma$-ray source was also found by the CGRO/EGRET (Paredes et al. 2000) and HESS (Aharonian et al. 2005) surveys. 
The observed features make LS~5039 a prominent member of the so-called gamma-ray binaries.

The optical companion of the system is a bright (V = 11.2), O6.5V((f)) type star classified by Clark et al. (2001) and 
McSwain et al. (2001, 2004). McSwain et al. (hereafter M04) presented the 
first orbital parameters of the binary system. Casares et al. (2005; hereafter C05) performed a detailed spectroscopic 
analysis of LS~5039, and calculated a new set of orbital and physical parameters. They found an orbital period of 3.906 
days, which has been confirmed by other studies: for example, variability also in X-rays (Bosch-Ramon et 
al. 2005, Takahashi et al. 2009), at GeV (Abdo et al. 2009) and TeV energies (Aharonian et al. 2006) occurs with the same 
period, suggesting orbital modulation of the high-energy emissions. 
Very recently, Aragona et al. (2009; hereafter A09) -- using the old and some new RV points -- also confirm the value of 
the orbital period and refined the parameters.

We have organised a simultaneous campaign of MOST space photometry and ground-based
optical spectroscopy, which was executed in July 2009. The main goal of the project was an investigation of the system
parameters (based on a new, independent, high-resolution data set), as well as an attempt to detect clumpiness in the 
stellar wind of the O-type star. Here we report on the first results of our spectroscopic analysis, while the detailed 
discussion of the results of the coordinated campaign will be presented elsewhere (Sarty et al., in prep).

\section{Observations and data reduction}

Spectroscopic observations were carried out on four nights between 2009 July 8-11 (parallel to MOST observations) and
on three nights between August 
1-3, using the 2.3-m telescope of the Australian National University (ANU) equipped with an echelle spectrograph. In 
total, the obtained 118 spectra cover almost 40 hours and ensure a good sampling of the whole orbit. 
The integration times were between 900-1200 s, while the spectra covered the whole visual
range between 3900 \AA\ and 6720 \AA\ . The nominal spectral resolution is $\lambda/\Delta \lambda\approx$ 23 000 at the 
H$\alpha$ line, with typical SNR of about 100 in 1 hour integration.

All data were reduced with standard IRAF\footnote{IRAF is distributed by the National Optical Astronomy 
Observatories, which are operated by the Association of Universities for Research in Astronomy, Inc., under cooperative 
agreement with the National Science Foundation.} tasks, including bias and flat-field corrections, cosmic ray removal, 
extraction of the 27 individual orders of the echelle spectra, wavelength calibration, and continuum normalization.

\section{Analysis and results}

\subsection{Radial velocities}

To measure radial velocities we first generated one hour long average spectra to get higher S/N -- one hour corresponds
to 0.01 orbital phase, hence negligible phase smearing appears in the phased radial velocity data.
The data were phased with an orbital period of 3.906 days (C05) and our RV curve confirms that this is indeed the correct
value. 

To generate the final RV diagrams we used average velocities of H \begin{small}I\end{small} (H$\alpha$, H$\beta$, 
H$\gamma$, H$\delta$, $\lambda$3835), He \begin{small}I\end{small} ($\lambda$4471, $\lambda$5875) and He 
\begin{small}II\end{small} ($\lambda$4200, $\lambda$4686, $\lambda$5411) lines; there are several other H and He lines in the
wavelength region of our spectra, but they were too noisy or blended to use them for velocity determination.
The typical measurement error is $\pm$10-15 km s$^{-1}$, which is partly caused by the observational noise, partly by the
large rotational velocity of the O-type star ($v$ sin $i$ = 113 $\pm$ 8 km s$^{-1}$, C05).

Our results indicate that there is a large systematic shift between He \begin{small}II\end{small}, He 
\begin{small}I\end{small} and H Balmer lines: the H \begin{small}I\end{small} and He \begin{small}I\end{small} lines are
always blueshifted by about 20 km s$^{-1}$ in respect to the He \begin{small}II\end{small} lines, which is a characteristic
signature of a strong stellar wind.
The photosphere of the O-type star is too hot for the H \begin{small}I\end{small} and He \begin{small}I\end{small} lines 
($\approx$ 40 000 K), hence these are generated in the cooler outer regions.
Interestingly, this velocity difference between the different species and ionisation states has not always been taken into
account: C05 also examined this effect (but they found smaller differences between velocities), but it was simply 
neglected by A09.
 
\begin{figure}[!ht]
\begin{center}
\begin{minipage}{15cm}
\includegraphics[width=7cm]{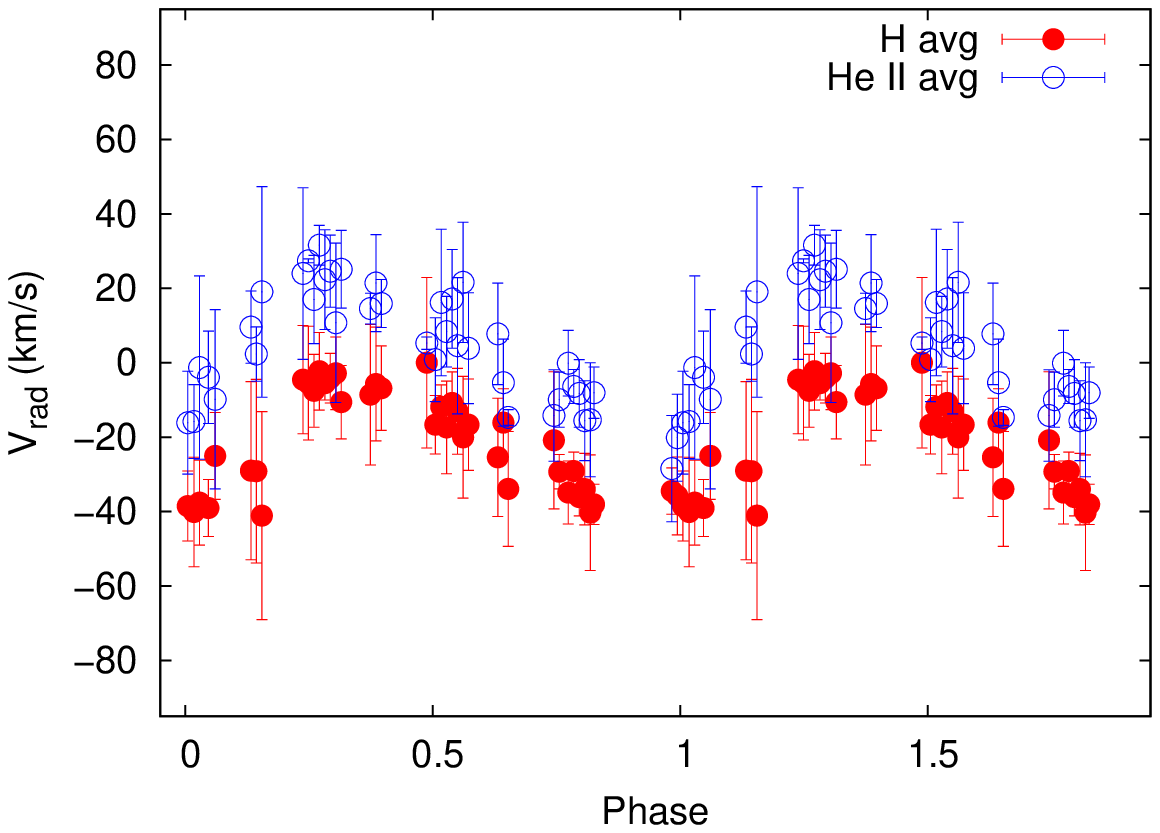}
\hspace{2mm}
\includegraphics[width=7cm]{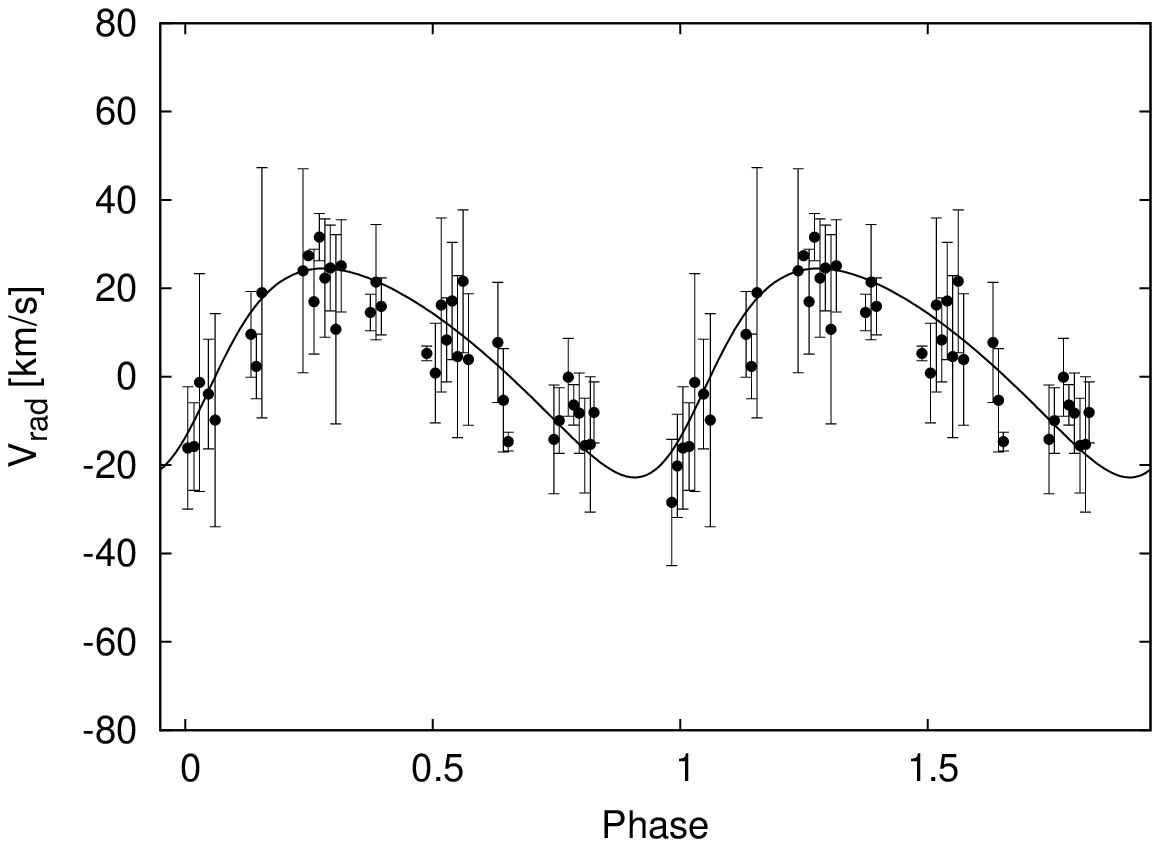}
\end{minipage}
\caption{{\it{Left}}: Radial velocities based on H Balmer and He II lines. {\it{Right}}: The best-fitting curve to radial 
velocities of He II lines.}
\label{rvfit}
\end{center}
\end{figure}

We also measured equivalent widths of several interstellar lines to estimate the rate
of interstellar reddening (see Section 3.2). During that we found redshifted 
satellite absorptions in the Ca \begin{small}II\end{small} K and Na \begin{small}I\end{small} D1\& D2 lines with a radial 
velocity around +60 km/s (more precisely, +58.4 $\pm$ 2.2 km s$^{-1}$ by Ca K, +62.9 $\pm$ 2.3 km s$^{-1}$ by Na D1 and 
+61.9 $\pm$ 1.7 km s$^{-1}$ by Na D2). These satellite lines may belong to a formerly unknown galactic Intermediate 
Velocity Cloud (IVC; see a review in Wakker \& van Woerden 1997).

\subsection{Orbital and physical parameters}

In the following analysis we adopted $T_{\rm eff}$ = 39 000 $\pm$ 1000 K and log $g$ = 3.85 $\pm$ 0.10 for the stellar 
companion (C05). To determine interstellar reddening we used Na \begin{small}I\end{small} 
D1 (${\it EW}$ = 0.70 $\pm$ 0.02 \AA\ ; see for ref. Munari \& Zwitter 1997), DIB $\lambda$5780 and DIB $\lambda$6613 
(${\it EW}$ = 0.55 $\pm$ 0.05 \AA\ and 0.18 $\pm$ 0.02 \AA\ , respectively, see for ref. Cox et al. 2005, Fig. 4.) lines, 
and we have got $E(B-V)$ = 1.2 $\pm$ 0.1, which is in very good agreement with previous results (Rib\'o et al. 2002, M04). 
Using other photometric parameters ($R$, $M_v$, see C05 and references therein) we also confirmed the distance ($d$ = 
2.5 $\pm$ 0.1 kpc), stellar radius ($R_0$ = 9.3$^{+0.7}_{-0.6}$ R$_{\odot}$) and mass of the O star ($M_0$ = 
22.9$^{+3.4}_{-2.9}$ M$_{\odot}$) published by C05.

Radial velocity curves were modelled using the 2003 version\footnote{ftp://ftp.astro.ufl.edu/pub/wilson} of the 
Wilson-Devinney (WD) code (Wilson \& Devinney 1971, Wilson \& van Hamme 2003). We note that during the analysis we only
used our own velocity points, which are independent from any previous data in the literature, and constitute the highest 
resolution, homogenous spectral dataset ever used to get the orbital solution of LS~5039.

The determined orbital parameters are shown in Table\ \ref{orbitpar}, compared 
to results of C05 and A09. The ones published by C05 were based only on velocity points related to 
He \begin{small}II\end{small} lines, while A09 combined the velocities of every available H, He \begin{small}I\end{small}, 
and He \begin{small}II\end{small} lines (see our cautionary notes above). 

In general, our parameters are close to previous solutions, but there are some differences. The value of $V_{\gamma}$ is 
definetly higher in C05 than in A09 or in our paper. The possibility of a real change in this parameter during only some 
years is very small, so the reason of the difference could be effect from different data treatment in of C05. The other 
important difference is our smaller value of $e$, what we explain with the smaller scatter and better coverage of our 
dataset.
\newpage

\begin{table}[!ht]
\caption{Orbital parameters of LS 5039.}
\begin{center}
\begin{tabular}{|l|l|l||l|}
\hline
Parameter & C05 (He \begin{small}II\end{small})& A09 & This paper (He \begin{small}II\end{small})\\
\hline
\hline
$T_0$ (HJD$-$2450000) & 1943.09 $\pm$ 0.10 & 2825.99 $\pm$ 0.05 & 5017.08 $\pm$ 0.06\\ 
$P_{orb}$ (d) & 3.90603 & 3.90608 & 3.906 (adopted)\\
$e$ & 0.35 $\pm$ 0.04 & 0.34 $\pm$ 0.04 & 0.24 $\pm$ 0.08\\
$\omega$ [$^{\circ}$] & 225.8 $\pm$ 3.3 & 236.0 $\pm$ 5.8 & 237.3 $\pm$ 21.8\\
$V_{\gamma}$ [km s$^{-1}$] & 17.2 $\pm$ 0.7 & 4.0 $\pm$ 0.3 & 3.9 $\pm$ 1.3\\
$K_1$ [km s$^{-1}$] & 25.2 $\pm$ 1.4 & 19.7 $\pm$ 0.9 & 23.6 $\pm$ 4.0\\
$a_1$ sin $i$ [R$_{\odot}$] & 1.82 $\pm$ 0.10 & 1.44 $\pm$ 0.07 & 1.77 $\pm$ 0.15\\
$f(m)$ [M$_{\odot}$] & 0.0053 $\pm$ 0.0009 & 0.0026 $\pm$ 0.0004 & 0.0049 $\pm$ 0.0006\\
rms of fit [km s$^{-1}$] & 9.1 & 7.1 & 6.2\\
\hline
\end{tabular}
\end{center}
\label{orbitpar}
\end{table}

\subsection{Circumstellar matter}

To infer mass loss rate of the O star and the properties of the circumstellar matter, we determined the 
equivalent widths of each H and He line and also their variability during the orbit, which could be a good indicator of the 
physical processes taking place in the stellar wind. We executed the measurements on the average 
spectra summarized from one hour long exposure times, but we plotted the daily average values of {\it EW}s on final diagrams 
(to better see the trends). 

For the H$\alpha$ line we found that {\it EW} changes between 2.50 and 2.85 \AA\ , and its variability shows a weak 
correlation with the orbital period. The main averege value (2.70 $\pm$ 0.12 \AA\ ) agree within uncertainties with the 
result of C05 (2.8 $\pm$ 0.1 \AA\ ), except that they found that value is stable during the orbit of the binary (which 
reason could be the relatively low resolution of their spectra). Our result is also consistent with the {\it EW} values 
measured in last ten years (Bosch-Ramon et al. 2007). Using the results of Puls et al. (1996, and references therein) we 
could estimate the mass loss rate from {\it EW} of H$\alpha$ line. To do these calculations we also had to use some other 
parameters: $R_0$ = 9.3$^{+0.7}_{-0.6}$ R$_{\odot}$ and $T_{\rm eff}$ = 39 000 $\pm$ 1000 K of the O star (C05), terminal 
velocity ($V_{\infty}$ = 2440 $\pm$ 190 km s$^{-1}$, M04) and wind velocity law exponent ($\beta$ = 0.8, M04).
We found that the mass-loss rate of the stellar companion is around 3.7 $\times$ 10$^{-7}$ M$_{\odot}$ yr$^{-1}$ by largest 
absorption (which correspond with the lower limit), while we could give 4.8 $\times$ 10$^{-7}$ M$_{\odot}$ yr$^{-1}$ for 
upper limit. These values also confirm the results of C05.

\begin{figure}[!ht]
\begin{center}
\begin{minipage}{15 cm}
\includegraphics[width=7cm]{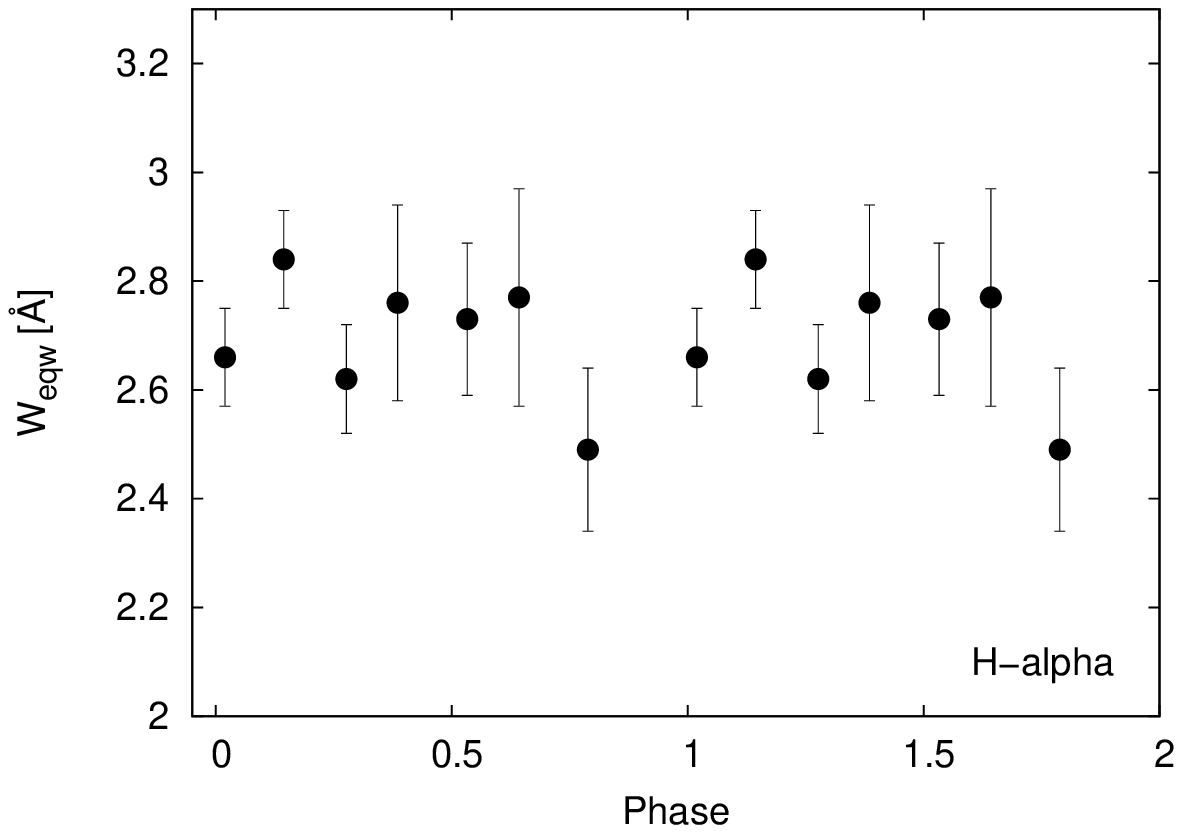}
\hspace{2mm}
\includegraphics[width=7cm]{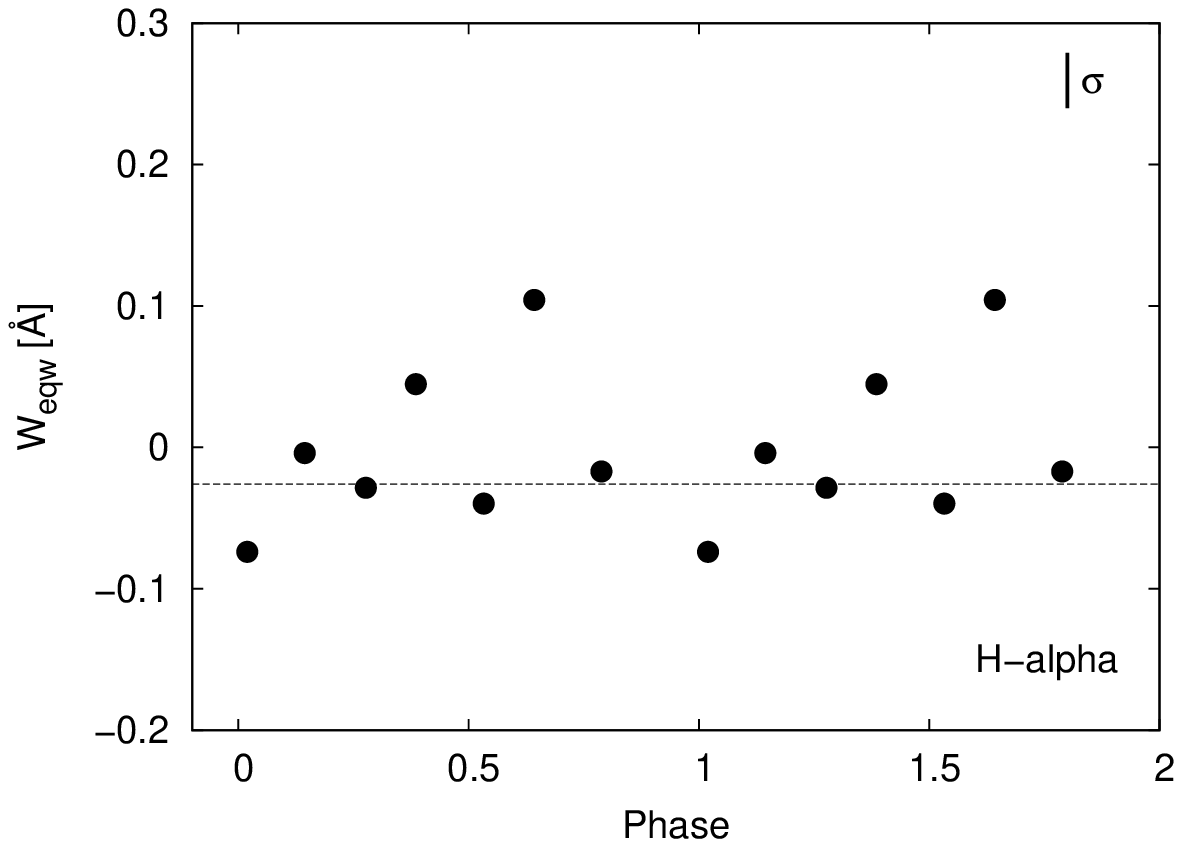}
\end{minipage}
\begin{minipage}{15 cm}
\includegraphics[width=7cm]{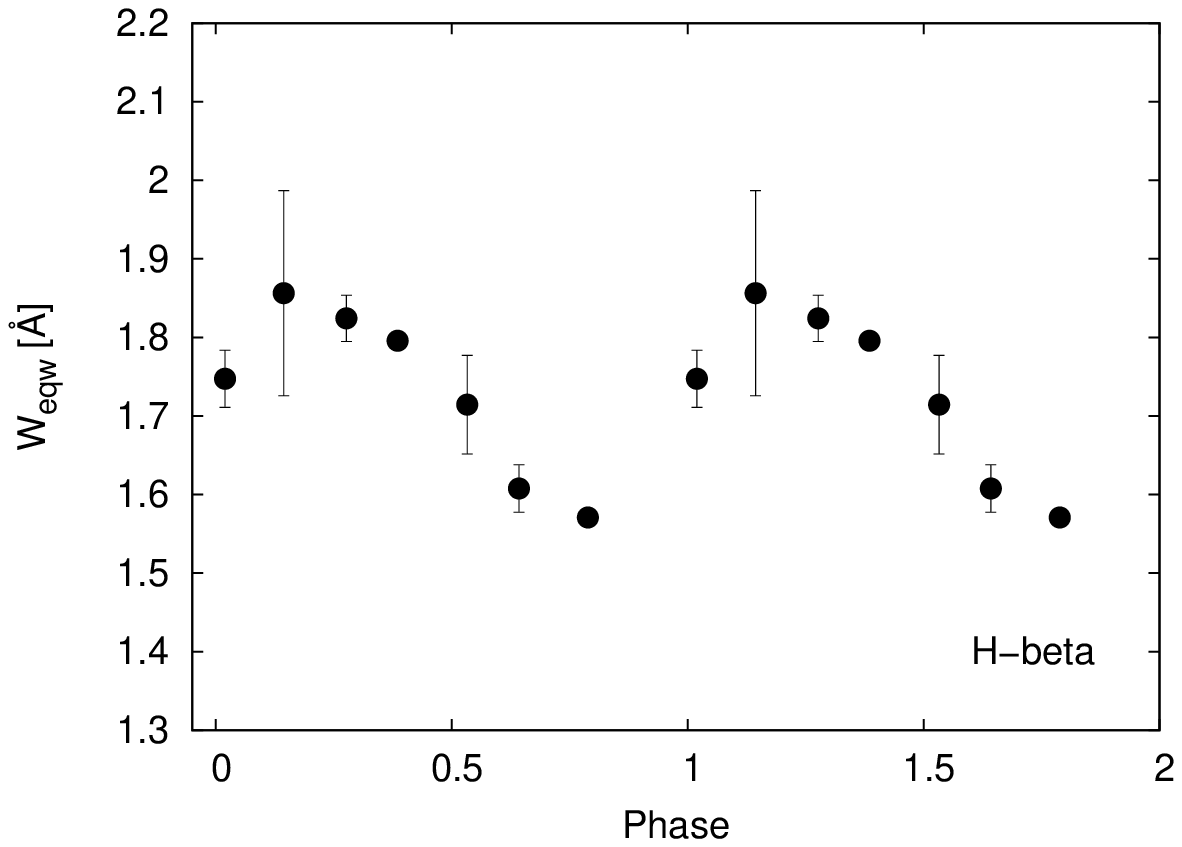}
\hspace{2mm}
\includegraphics[width=7cm]{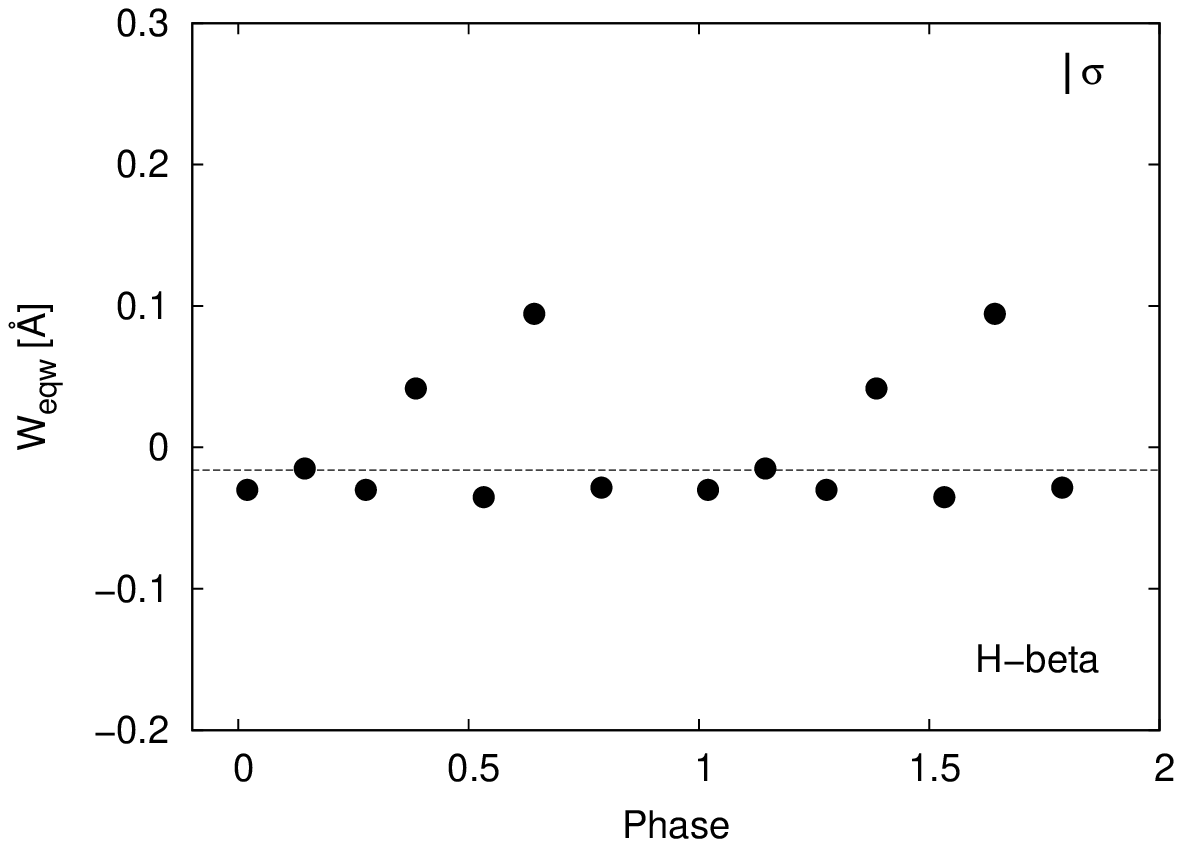}
\end{minipage}
\begin{minipage}{15 cm}
\includegraphics[width=7cm]{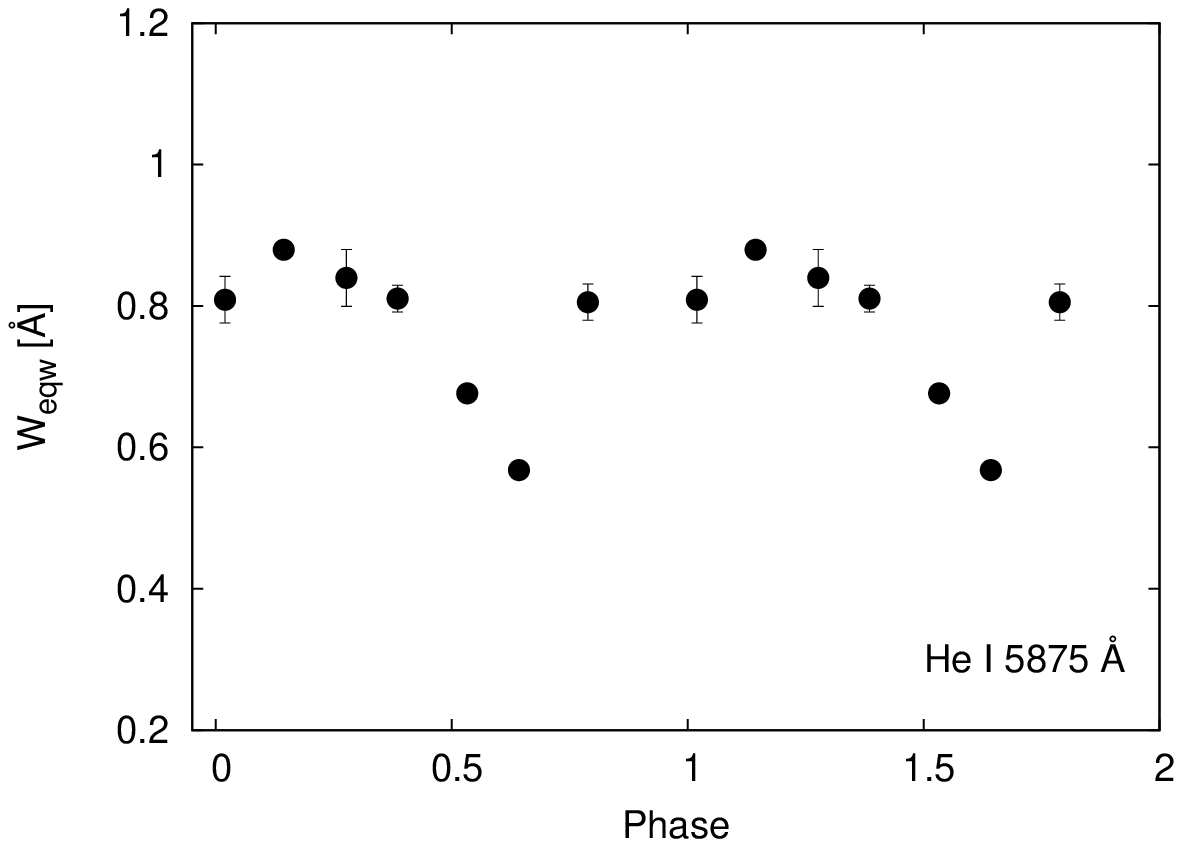}
\hspace{2mm}
\includegraphics[width=7cm]{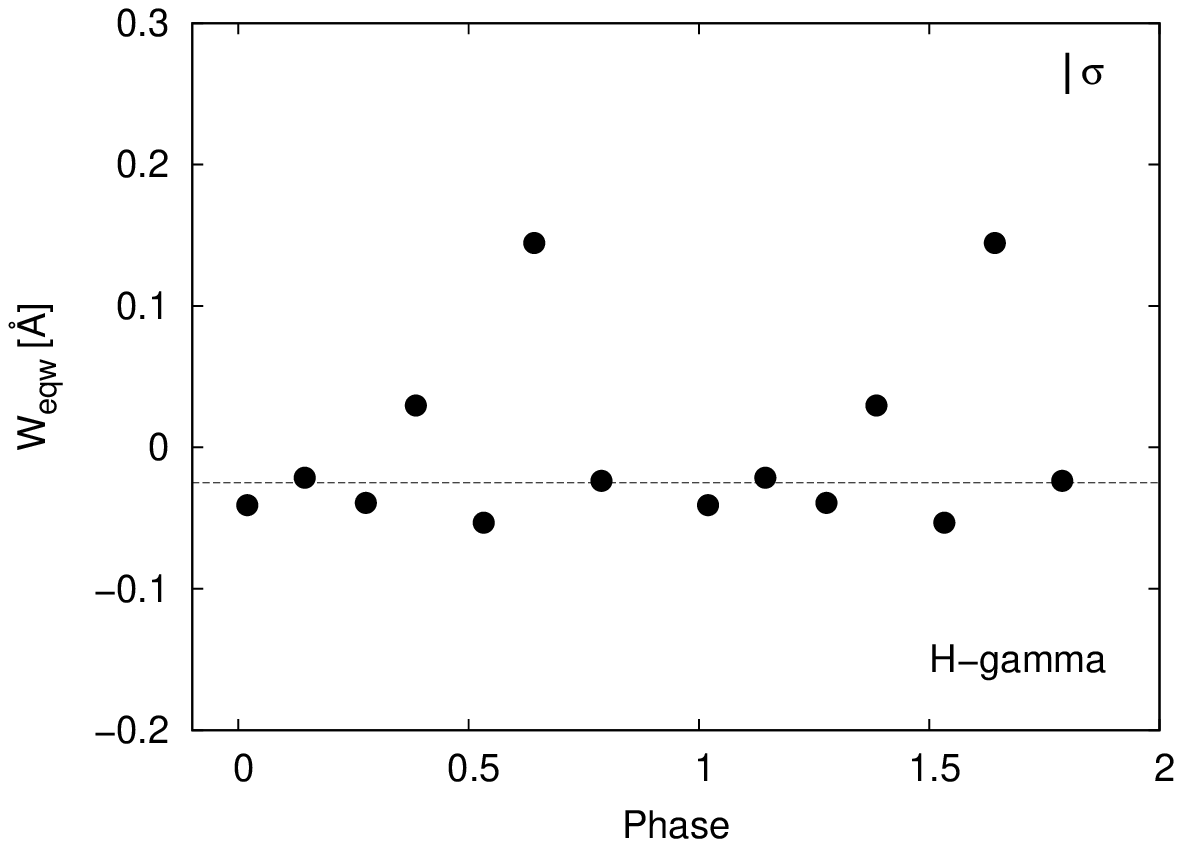}
\end{minipage}
\caption{{\it{Left}}: Variability of equivalent widths of H$\alpha$, H$\beta$ and He I $\lambda$5875 lines during the orbit.
{\it{Right}}: Variability of equivalent widths of the emission components of Balmer lines during the orbit of LS~5039. In 
these diagrams larger numbers correspond to larger emission rates. There is a point by each line with a significantly high 
value (3, 3 and 5 $\sigma$ above average, respectively) at $\varphi\sim$0.65.}
\label{abs_em_EW}
\end{center}
\end{figure}

There are two lines (H$\beta$ and He \begin{small}I\end{small} $\lambda$5875) by which the values of {\it EW} are more 
strongly modulated by orbital period. The lowest absorption is around $\varphi\sim$0.75 by H lines, and $\varphi\sim$0.65 by 
the mentioned He line (Fig.\ \ref{abs_em_EW}, left side), close to the expected phase $\varphi\sim$0.7 when the compact 
object is between us and the stellar companion (inferior conjunction).

By H Balmer lines we also tried to determine the phase dependence of {\it EW}s only of emission 
components. Because of the relatively small S/N it was not performable to follow hour-to-hour changes in line profiles, so
therefore we generated difference spectra subtracted Doppler-corrected daily average spectra from their main average to
calculate {\it EW}s of remaining emission components. Using these values quantitatively would be an excessive implementation of
results, but the relative, phase dependent changes are examinable.

The three diagrams (Fig.\ \ref{abs_em_EW}, right side) are very similar to each other, with a significantly exceeding point 
(which means maximal emission here) near inferior conjunction ($\varphi\sim$0.65). It agrees with the diagrams shown on the 
left side of Fig.\ \ref{abs_em_EW}, where lowest absorption is also at that phase. We can rule out any technical artifacts 
which could produce the similarity of the three diagrams, because the lines are in different echelle orders, so they went 
independently through the different steps of data reductions.

\section{Summary}

In our work we presented the first results of a coordinated campaign consisting of parallel space photometry and
high-resolution ground-based spectroscopy of the the X-ray (gamma-ray) binary LS~5039. Detailed analysis of spectroscopic 
data (which incarnate the highest resolution, homogenous dataset ever obtained from that object) allowed us to determine
the orbital parameters of the system being mainly consistent with former solutions, but we also found some differences 
(emphasizing the significant velocity shift between H \begin{small}I\end{small}, He \begin{small}I\end{small} and 
He \begin{small}II\end{small} lines, and the smaller value of eccentricity).

To reveal the properties of the circumstellar environment, we first measured the equivalent widhts of available H
and He lines during the orbit. We found that H$\alpha$, H$\beta$ and He \begin{small}I\end{small} $\lambda$5875 lines are
unambiguously modulated by orbital period -- the clear emission components of H Balmer lines show similar phase 
dependence, which seems to confirm that it is a real effect. From the average {\it EW} of H$\alpha$ line we estimated the 
mass loss rate of the O star and got similar value to former results.

The complete analysis of our observations about LS~5039 will be presented in a latter paper (Sarty et al., in prep),
including discussion about the circumstellar environment and possible mass ranges of the compact object.

\section*{Acknowledgments} 

This work has been supported by the Australian Research Council, the University
of Sydney, the Hungarian OTKA Grant K76816 and the ``Lend\"ulet'' Young
Researchers' Program of the Hungarian Academy of Sciences. 

\section*{References}

\begin{thereferences}

\item
  Abdo, A. A. et al. 2009, ApJ, 706, L56
\item
  Aharonian, F. A. et al. 2005, Science, 309, 746
\item
  Aharonian, F. A. et al. 2006, A\&A, 460, 743
\item
  Aragona, C., McSwain, M. V., Grundstrom, E. D., Marsh, A. N., Roettenbacher, R. M., Hessler, K. M., Boyajian, T. S., and Ray, P. S. 2009, ApJ, 698, 514 (A09)
\item
  Bosch-Ramon, V., Paredes, J. M., Rib\'o, M., Miller, J. M., Reig, P., and Mart\'i, J. 2005, ApJ, 628, 388
\item
  Bosch-Ramon, V., Motch, C., Rib\'o, M., Lopes de Oliveira, R., Janot-Pacheco, E., Negueruela, I., Paredes, J. M., and Martiocchia, A. 2007, A\&A, 473, 545  
\item
  Casares, J., Rib\'o, M., Ribas, I., Paredes, J. M., Mart\'i, J., and Herrero, A. 2005, MNRAS, 364, 899 (C05)
\item
  Clark, J. S., Reig, P., Goodwin, S. P., Larionov, V. M., Blay, P., Coe, M. J., Fabregat, J., Negueruela, I., Papadakis, I., and Steele, I. A. 2001, A\&A, 376, 476
\item
  Cox, N. L. J., Kaper, L., Foing, B. H., and Ehrenfreund, P. 2005, A\&A, 438, 187
\item
  Mart\'i, J., Paredes, J. M., and Rib\'o, M. 1998, A\&A, 338, L71
\item
  McSwain, M. V., Gies, D. R., Huang, W., Wiita, P. J., and Wingert, D. W. 2004, ApJ, 600, 927 (M04)
\item
  McSwain, M. V., Gies, D. R., Riddle, R. L., Wang, Z., and Wingert, D. W. 2001, ApJ, 558, L43
\item
  Motch, C., Haberl, F., Dennerl, K., Pakull, M., and Janot-Pacheco, E. 1997, A\&A, 323, 835
\item
  Munari, U., and Zwitter, T. 1997, A\&A, 318, 269
\item
  Paredes, J. M., Mart\'i, J., Rib\'o, M., and Massi, M. 2000, Science, 288, 2340 
\item
  Paredes, J. M., Rib\'o, M., Ros, E., Mart\'i, J., and Massi, M. 2002, A\&A, 393, L99
\item
  Puls, J., et al. 1996, A\&A, 305, 171 
\item
  Rib\'o, M., Paredes, J. M., Romero, G. E., Benaglia, P., Mart\'i, J., Fors, O., and Garc\'ia-S\'anchez, J. 2002, A\&A, 384, 954
\item
  Takahashi, T., Kishishita, T., Uchiyama, Y., Tanaka, T., Yamaoka, K., Khangulyan, D., Aharonian, F. A., Bosch-Ramon, V., and Hinton, J. A. 2009, ApJ, 697, 592
\item
  Wakker, B. P., and van Woerden, H. 1997, ARA\&A, 35, 217
\item
  Wilson, R. E., and Devinney, E. J. 1971, ApJ, 166, 605
\item
  Wilson, R. E., and van Hamme, W. 2003, Computing Binary Stars Observables, ver. 4 (Gainesville: University of Florida)

\end{thereferences}

\end{document}